\documentclass[reprint, twocolumn, nofootinbib,
nobibnotes, aps, prl, showpacs, floatfix]{revtex4-1}

\usepackage{dcolumn}
\usepackage{amsmath}
\usepackage{booktabs}
\usepackage{graphicx}
\usepackage{multirow}
\usepackage{lipsum}

\begin{document}

\title{Comparison of Hartree-Fock and Hartree-Fock-Slater approximations for calculation of radiation damage dynamics of light and heavy atoms in the field of a x-ray free electron laser.}
\author{A. Kozlov, H. M. Quiney}

\affiliation{ARC Centre of Excellence for Advanced Molecular Imaging}
\affiliation{School of Physics, The University of Melbourne, 
Victoria 3010, Australia}
\date{\today}

\begin{abstract}
Simulations of radiation damage in single molecule imaging using a X-ray free electron laser use atomic rates calculated in the lowest order. We investigate the difference in ion yield predictions using Hartree-Fock and Hartree-Fock-Slater approximations for light and heavy elements of biological significance. The results show that for the biologically abundant elements of the second and third rows of the periodic table both approximations agree to about 6\%. For the heavier elements beyond the fourth row the discrepancy rices to 11\% for the range of the pulse parameters covered in this work. Presented analysis can be used for an error estimation in a wide range of ab initio simulations of the X-ray pulse interaction with biological molecules. We also discuss other atomic structure effects and show that their account has considerably smaller effect on the ion yields of respective elements compared to the choice of the approximation.
\end{abstract}

\maketitle

\section{Introduction}
Single molecule imaging using a X-ray free electron laser has the potential to outrun radiation damage in the sample \cite{Neutze2000}. For typical intensities and pulse lengths of the order of ten femtoseconds radiation damage becomes dynamic and nonlinear which rapidly distorts the electronic configuration of the target \cite{C60}. However, studies suggest \cite{Quiney2011, Barty2013, Martin2015} that the impact of damage might have a limited effect on the reconstructed molecular structure compared to the noise and resolution in the diffraction data \cite{Gureyev2018}.

Accurate \textit{ab initio} simulation of XFEL interaction with large molecules is a challenging task due to the importance of both structure and dynamics. This goal is beyond any currently available computation resources, so that simplifications are made. The high velocities that particles acquire shortly after the start of the pulse allow the use of molecular dynamics methods which treat the problem of free electrons and ions motion as classical. For a molecule with a few atoms, the inclusion of molecular structure effects can be treated at various levels of theory \cite{Rudenko2017, Picon2017}. For large molecules the only currently available technique treats a molecule as a collection of atoms which evolve independently from one another, subject to the XFEL radiation pulse and free electrons \cite{Moribayashi2010, Martin2015} released by atoms. With these approximations one can use molecular dynamics \cite{Ostlin2017, Ho2017}  or plasma-type simulations \cite{Nakamura2009, Martin2015} with classical motion of atoms, ions, and free electrons and quantum dynamics for the states of individual ions.

For the independent atom model of biomolecules in XFEL pulses, various damage effects, such as Compton scattering \cite{Santra2014} and electron impact ionization \cite{Kai2009, Gorobtsov2015}, have been studied. In this paper we compare the predictions of the Hartree-Fock and Hartree-Fock-Slater methods at the  level of theory which is typically used. We investigate the discrepancy between the ion yields predicted by both models. The corresponding charge state dynamics is typically used in a large scale simulations of XFEL pulse interaction with biological molecules. We limit our considerations to the X-ray photon energy of 8 keV and vary the XFEL pulse width from 2 to 50 femtoseconds, as well as the peak pulse intensity from $2.5\times 10^{18}$ to $5\times 10^{20}$ W/cm$^2$. We observe that for the elements in the second and third rows of the periodic table the discrepancy in ion yields can be as large as 6\%, but for iron atom it reaches 11\% for weaker and shorter pulses. When the pulse energy is increased the results produced by both approximations converge. Other factors, such as the choice of length and velocity gauges for photoionization matrix elements in the Hartree-Fock approximation, or the relativistic effects are shown to have a much smaller effect on the ion yields and the charge state dynamics of the considered elements.

\section{Mean field atomic structure calculations}
Atomic structure theory has been used for the evaluation of rates for different processes since its foundation by Hartree \cite{Hartree1928} and Fock \cite{Fock1930}. Numerous improvements (see for example \cite{JohnsonBook, GrantBook}) were developed and allow accurate calculations to be carried out. Even a molecular mean field calculations are not used for time-dependent simulation of the scattering and radiation damage processes in large biological molecules that typically contain thousands of atoms due to current computational limitations. For smaller molecules such calculations have recently been published $[7, 8]$. For large biological molecules and atomic clusters a combination of mean field atomic calculations for radiation damage rates coupled with molecular dynamics simulation for atomic motion remains the only currently available option \cite{Ostlin2017, Ho2017}. This approach is often refereed to as an independent atom model. Experimetal evidence \cite{Davis1993} suggests that for small molecules ionization cross-sections for soft X-rays are well approximated by a sum of atomic cross-sections, while molecular Auger rates agree to 25\% with the respective atomic values \cite{Thomas1991}. A more detailed discussion on independent atom model for ionization dynamics can be found in \cite{Moribayashi2008}. In this model decay rates and ionization cross-sections are estimated using an atomic calculations, which in turn determine the X-ray scattering via a time-dependent formfactor \cite{Santra2011} and molecular dynamics through a time-dependent ionic charge. In this work we address how the different choice of exchange potential in atomic calculations affects the time-dependent charge state dynamics for all possible ionic states of initial atoms.

The time-independent Schr\"odinger equation for a single electron orbital \cite{JohnsonBook} in the field of the nucleus and other electrons can be written as
\begin{equation}\label{HFequation}
\left( -\frac{1}{2}\Delta - \frac{Z}{r} + \hat V_{dir}(\vec{r}\,) + \hat V_{ex}(\vec{r}\,) \right)\psi_a(\vec{r}\,) = \epsilon_a \psi_a(\vec{r}\,),
\end{equation}
where $\psi_a(\vec{r})$ is the bound state wavefunction with energy $\epsilon_a$ and $\hat V_{dir}(\vec{r}\,)$ and $\hat V_{ex}(\vec{r}\,)$ are direct and exchange part of the mean field potential. The above equation is solved self consistently for all occupied bound state orbitals of a given atom or ion. The direct part of the potential has a classical analog and represents electrostatic repulsion between the electron in orbital $a$ and all other electrons:
\begin{equation}
\hat V_{dir}(\vec{r}\,) = \sum_b \langle b | \frac{1}{|\vec{r} - \vec{r}\,'|}| b \rangle
\end{equation}
Here and below the Dirac notations are used for brevity, so that $\langle b | \hat A(\vec{r}\,') | b \rangle = \int d\vec{r}\,' \psi_b^*(\vec{r}\,') \hat A(\vec{r}\,') \psi_b(\vec{r}\,')$ for a matrix element of an operator $\hat A(\vec{r}\,')$. Unlike the direct part, the exchange potential can be described using several different forms. The Hartree-Fock-Slater (HFS) approximation is widely used for modeling of atom interaction with XFEL pulse \cite{Santra2011, Moribayashi2008, Ho2017, Rudenko2017}. The exchange potential in this case is local and is given by
\begin{equation}
\hat V_{ex}(\vec{r}\,) = -\frac{3}{2}\left(\frac{3}{\pi} \sum_b | \psi_b(\vec{r}\,)|^2 \right)^{1/3},
\end{equation}
where $b$ indicates the summation over all bound state orbitals. The advantage of the HFS approximation is its computation simplicity and ease of  implementation. This comes at a price of accuracy which is rather poor for HFS, especially for a valence shell orbitals. A more accurate expression for the exchange potential is obtained in the Hartree-Fock (HF) approximation \cite{Hartree1928, Fock1930}. Unlike HFS it is non-local and is defined through its action on the occupied wavefunction
\begin{equation}
\hat V_{ex}(\vec{r}\,) \psi_a(\vec{r}\,) = -\sum_b\psi_b(\vec{r}\,) \langle b | \frac{1}{|\vec{r} - \vec{r}\,'|}| a \rangle.
\end{equation}
As a result it is more difficult to implement and the computational cost is higher compared to HFS.

We use the restricted average over configuration form of equation (\ref{HFequation}). A detailed description can be found in many textbooks on atomic physics, for example \cite{GrantBook}. For evaluation of the excited states (including the continuum states) of an atom the potential needs to be adjusted to account for the reduction in the number of bound state electrons. Since the effective potentials for initial and final states are different in this approach the length and velocity gauges no longer result in the same radiative matrix elements \cite{ParkBook}. In the following section we consider continuum states in more detail. 

\begin{table*}\center
\caption{Comparison of photoionization cross-sections $\sigma_P$  for atomic carbon calculated using Hartree-Fock and Hartree-Fock-Slater (local form of exchange) potentials. Photon energy is 8 keV, cross-sections are given in $10^{-8}$ a.u. The values in brackets were obtained in \cite{Santra2011} with Hartree-Fock-Slater potential. } 
{\renewcommand{\arraystretch}{0}%
\begin{tabular}{c c c c c c c c c c}
\hline\hline
\rule{0pt}{4pt}\\
charge & configuration &  & \multicolumn{3}{c}{$V^{N-1}$, HF } & & \multicolumn{3}{c}{$V^{N}$, HFS(\cite{Santra2011})} \\ 
\rule{0pt}{2pt}\\
& & & $1s$ & $2s$ & $2p$ & & $1s$ & $2s$ & $2p$ \\ 
\rule{0pt}{4pt}\\
\hline
\rule{0pt}{4pt}\\
0 & $1s^22s^22p^2$ &  & 265 & 11.0 & 0.0575 & & 287(287) & 14.8(14.8) & 0.0897(0.0897)\\
\rule{0pt}{2pt}\\
\hline
\rule{0pt}{2pt}\\
+1 & $1s^12s^22p^2$ &  & 152 & 16.3 & 0.132 & & 155(155) & 20.9(20.8) & 0.218(0.219)\\
\rule{0pt}{2pt}\\
 & $1s^22s^12p^2$ &  & 265 & 6.51 & 0.0736 & & 287(287) & 8.29(8.28) & 0.117(0.118)\\
\rule{0pt}{2pt}\\
 & $1s^22s^22p^1$ &  & 266 & 1.26 & 0.0373 & & 287(287) & 16.6(16.6) & 0.0588(0.0590)\\
\rule{0pt}{4pt}\\
\hline
\rule{0pt}{2pt}\\
+2 & $1s^02s^22p^2$ &  & - & 26.6 & 0.344 & & - & 27.9(27.9) & 0.387(0.387)\\
\rule{0pt}{2pt}\\
 & $1s^22s^02p^2$ &  & 265 & - & 0.0949 & & 288(288) & - & 0.145(0.145)\\
\rule{0pt}{2pt}\\
 & $1s^22s^22p^0$ &  & 267 & 14.8 & - & & 288(288) & 18.7(18.6) & -\\
\rule{0pt}{2pt}\\
 & $1s^12s^12p^2$ &  & 152 & 9.4 & 0.163 & & 155(155) & 11.6(11.5) & 0.257(0.258)\\
\rule{0pt}{2pt}\\
 & $1s^12s^22p^1$ &  & 152 & 18.6 & 0.0844 & & 155(155) & 23.5(23.4) & 0.132(0.132)\\
\rule{0pt}{2pt}\\
 & $1s^22s^12p^1$ &  & 266 & 7.51 & 0.0485 & & 288(288) & 9.27(9.25) & 0.0735(0.0737)\\
\rule{0pt}{4pt}\\
\hline
\rule{0pt}{2pt}\\
+3 & $1s^02s^12p^2$ &  & - & 15.0 & 0.400 & & - & 15.1(15.1) & 0.436(0.435)\\
\rule{0pt}{2pt}\\
 & $1s^02s^22p^1$ &  & - & 29.9 & 0.208 & & - & 31.1(31.1) & 0.221(0.221)\\
\rule{0pt}{2pt}\\
 & $1s^12s^02p^2$ &  & 152 & - & 0.201 & & 155(156) & - & 0.300(0.300)\\
\rule{0pt}{2pt}\\
 & $1s^22s^02p^1$ &  & 266 & - & 0.0629 & & 289(289) & - & 0.0884(0.0885)\\
\rule{0pt}{2pt}\\
 & $1s^12s^22p^0$ &  & 153 & 21.3 & - & & 156(156) & 26.0(26.0) & -\\
\rule{0pt}{2pt}\\
 & $1s^22s^12p^0$ &  & 267 & 8.74 & - & & 289(289) & 10.8(10.8) & -\\
 \rule{0pt}{2pt}\\
 & $1s^12s^12p^1$ &  & 152 & 10.7 & 0.103 & & 156(156) & 12.8(12.8) & 0.153(0.153)\\
 \hline
\rule{0pt}{2pt}\\
+4 & $1s^02s^02p^2$ &  & - & - & 0.477 & & - & - & 0.476(0.454)\\
\rule{0pt}{2pt}\\
 & $1s^02s^22p^0$ &  & - & 33.4 & - & & - & 36.9(33.8) & -\\
 \rule{0pt}{2pt}\\
 & $1s^22s^02p^0$ &  & 268 & - & - & & 254(284) & - & -\\
\rule{0pt}{2pt}\\
 & $1s^02s^12p^1$ &  & - & 16.7 & 0.241 & & - & 16.7(16.6) & 0.246(0.246)\\
\rule{0pt}{2pt}\\
 & $1s^12s^12p^0$ &  & 153 & 12.2 & - & & 157(157) & 14.7(14.7) & -\\
\rule{0pt}{2pt}\\
 & $1s^12s^02p^1$ &  & 152 & - & 0.126 & & 156(156) & - & 0.180(0.180)\\
  \hline
\rule{0pt}{2pt}\\
+5 & $1s^12s^02p^0$ &  & 156 & - & - & & 156(156) & - & -\\
\rule{0pt}{2pt}\\
 & $1s^02s^12p^0$ &  & - & 18.7 & - & & - & 18.7(18.0) & -\\
 \rule{0pt}{2pt}\\
 & $1s^02s^02p^1$ &  & - & - & 0.286 & & - & - & 0.286(0.285)\\
\rule{0pt}{4pt}\\
\hline\hline
\end{tabular}}\label{PhotoTab}
\end{table*}

\begin{table*}\center
\caption{Photoionization cross-sections of carbon atom in length and velocity gauges in $V^{N-1}$  (HFS $V^N$ added for comparison). Photon energy is 8 keV, cross-sections are given in $10^{-8}$ a.u.} 
{\renewcommand{\arraystretch}{0}%
\begin{tabular}{c c c c c c c c c c}
\hline\hline
\rule{0pt}{4pt}\\
charge & configuration &  & \multicolumn{3}{c}{$V^{N-1}$, HF length(velocity) gauge} & & \multicolumn{3}{c}{$V^N$, HFS} \\ 
\rule{0pt}{2pt}\\
& & & $1s$ & $2s$ & $2p$ & & $1s$ & $2s$ & $2p$ \\ 
\rule{0pt}{4pt}\\
\hline
\rule{0pt}{4pt}\\
0 & $1s^22s^22p^2$ &  & 265(272) & 11.0(12.0) & 0.0575(0.0636) & & 287 & 14.8 & 0.0897\\
\rule{0pt}{2pt}\\
\hline
\rule{0pt}{2pt}\\
+1 & $1s^12s^22p^2$ &  & 152(154) & 16.3(18.2) & 0.132(0.166) & & 155 & 20.9 & 0.218\\
\rule{0pt}{2pt}\\
 & $1s^22s^12p^2$ &  & 265(273) & 6.51(7.06) & 0.0736(0.0870) & & 287 & 8.29 & 0.117\\
\rule{0pt}{2pt}\\
 & $1s^22s^22p^1$ &  & 266(273) & 12.6(14.2) & 0.0373(0.0443) & & 287 & 16.6 & 0.0588\\
\rule{0pt}{4pt}\\
\hline
\rule{0pt}{2pt}\\
+2 & $1s^02s^22p^2$ &  & - & 26.6(26.9) & 0.344(0.352) & & - & 27.9 & 0.387\\
\rule{0pt}{2pt}\\
 & $1s^22s^02p^2$ &  & 265(273) & - & 0.0949(0.115) & & 288 & - &0.145\\
\rule{0pt}{2pt}\\
 & $1s^22s^22p^0$ &  & 267(274) & 14.8(16.0) & - & & 288 & 18.7 & -\\
\rule{0pt}{2pt}\\
 & $1s^12s^12p^2$ &  & 152(154) & 9.4(10.5) & 0.163(0.204) & & 155 & 11.6 & 0.257\\
\rule{0pt}{2pt}\\
 & $1s^12s^22p^1$ &  & 152(154) & 18.6(20.7) & 0.0844(0.106) & & 155 & 23.5 & 0.132\\
\rule{0pt}{2pt}\\
 & $1s^22s^12p^1$ &  & 266(273) & 7.51(8.1) & 0.0485(0.0585) & & 288 & 9.27 & 0.0735\\
\rule{0pt}{4pt}\\
\hline
\rule{0pt}{2pt}\\
+3 & $1s^02s^12p^2$ &  & - & 15.0(15.1) & 0.400(0.410) & & - & 15.1 & 0.436\\
\rule{0pt}{2pt}\\
 & $1s^02s^22p^1$ &  & - & 29.9(30.2) & 0.208(0.212) & & - & 31.1 & 0.221\\
\rule{0pt}{2pt}\\
 & $1s^12s^02p^2$ &  & 152(154) & - & 0.200(0.250) & & 156 & - & 0.300\\
\rule{0pt}{2pt}\\
 & $1s^22s^02p^1$ &  & 266(274) & - & 0.0629(0.0755) & & 289 & - & 0.0884\\
\rule{0pt}{2pt}\\
 & $1s^12s^22p^0$ &  & 153(155) & 21.3(23.6) & - & & 156 & 26.0 & -\\
\rule{0pt}{2pt}\\
 & $1s^22s^12p^0$ &  & 267(274) & 8.7(9.4) & - & & 289 & 10.8 & -\\
 \rule{0pt}{2pt}\\
 & $1s^12s^12p^1$ &  & 152(154) & 10.7(11.9) & 0.103(0.128) & & 156 & 12.8 & 0.153\\
 \hline
\rule{0pt}{2pt}\\
+4 & $1s^02s^02p^2$ &  & - & - & 0.477(0.477) & & - & - & 0.369\\
\rule{0pt}{2pt}\\
 & $1s^02s^22p^0$ &  & - & 33.4(33.8) & - & & - & 25.4 & -\\
 \rule{0pt}{2pt}\\
 & $1s^22s^02p^0$ &  & 268(275) & - & - & & 290 & - & -\\
\rule{0pt}{2pt}\\
 & $1s^02s^12p^1$ &  & - & 16.7(16.9) & 0.241(0.246) & & - & 16.7 & 0.246\\
\rule{0pt}{2pt}\\
 & $1s^12s^12p^0$ &  & 153(155) & 12.2(13.4) & - & & 157 & 14.7 & -\\
\rule{0pt}{2pt}\\
& $1s^12s^02p^1$ &  & 152(154) & - & 0.126(0.154) & & 156 & - & 0.180\\
  \hline
\rule{0pt}{2pt}\\
+5 & $1s^12s^02p^0$ &  & 156(156) & - & - & & 156 & - & -\\
\rule{0pt}{2pt}\\
 & $1s^02s^12p^0$ &  & - & 18.7(18.7) & - & & - & 18.7 & -\\
 \rule{0pt}{2pt}\\
 & $1s^02s^02p^1$ &  & - & - & 0.286(0.286) & & - & - & 0.286\\
\rule{0pt}{4pt}\\
\hline\hline
\end{tabular}}\label{PhotoGaugeTab}
\end{table*}

\section{Length and velocity gauge for photoionization cross-sections}
The absorption of photons by an atom leads to photoionization, provided that the photon energy exceeds the ionization potential. In the Hartree-Fock approximation the ionization potential is given by $I = -\epsilon_i$, according to Koopmans theorem \cite{Koopmans1934}, therefore the final continuum state energy of the escaping electron is $\epsilon_c = \omega + \epsilon_i$. For HFS orbitals this expression is only approximately correct. We consider photoionization process in the electric dipole approximation so the corresponding selection rules should be applied. In the non-relativistic case it means that the final and initial orbitals should have opposite parity, conserved electron spin direction and the angular momentum change restricted to $|l_i - l_c| = 1$.

Total photoionization cross-section in the length gauge \cite{ParkBook} averaged over initial state and summed over final states can be written as 
\begin{equation}\label{Photo}
\sigma_P = \frac{4}{3}\alpha \pi^2 \omega N_e\sum_{l_c = |l_i-1|}^{l_i+1} \frac{l_>}{2l_i+1}\left|\int_0^{\infty} drP_{i}(r)rP_{c}(r)\right|^2,
\end{equation}
where $\alpha$ is the fine structure constant, $N_e$ is the number of electrons in orbital $i$, $l_> = \max(l_i, l_c)$. Calculation of radial continuum wavefunctions $P_c(r)$ in (\ref{Photo}) becomes progressively more demanding with increasing photon energy. In the case of X-ray photons the initial orbital $P_e(r)$ is a slowly varying function compared to $P_c(r)$ so that the integral in (\ref{Photo}) becomes small with increasing $\epsilon_i$. In this case replacement of $P_c(r)$ with its asymptotic expression may result in a large error in the matrix element value. In this case one should numerically integrate $P_c(r)$ to the end of the numerical mesh or to a practical infinity of $P_i(r)$ provided that the asymptotic limit is reached. We chose a log-linear mesh \cite{Amusia1997} which is uniform in $\rho = r + \beta \ln r$ with small value of parameter $\beta\approx10^{-3}$ to insure that there are at least 40 points per wavelength in the asymptotic region. 

As mentioned in the previous section, the mean field potential for evaluation of $P_c(r)$ needs to be chosen. For an \textit{ ab initio} calculation it is natural to use the $V^{N-1}$ HF potential \cite{JohnsonBook, Dzuba2005} since it accounts for the reduction in the number of remaining electrons and insures orthogonality of the core and the excited states. Although in some cases relaxing the orthogonality restriction provides a better initial approximation for continuum states \cite{Kelly1975}, the overlap integral of $P_c(r)$ and the bound orbitals becomes vanishingly small for energies above a few atomic units \cite{Kozlov2017}. Table \ref{PhotoTab} allows the comparison of our results for cross-sections with the existing calculations for carbon atom. The agreement is reasonable and within $10\%$ for inner shell orbitals, educing to about $30\%$ for outer shells. We observe a similar pattern for all other elements for which we carried out calculations. The reason for the discrepancy between the HFS and HF results is intuitively clear. As one progresses from the inner to valence shells, the two approaches give different screening of the nuclear potential. We accurately reproduced the results of \cite{Santra2011} using the $V^N$ HFS approximation with the Latter tail correction, in which the excited states are calculated in the field of all initial electrons. Such a potential corresponds to $N+1$ particles in the final state of the system which results in rather a poor approximation for excited states without further application of post-Hartree-Fock methods. Comparison of our result ($74.19$ barn for $V^{N-1}$ HF and $80.35$ barn for $V^N$ HFS potentials) with the relativistic $V^N$ HFS calculations of ref. \cite{Scofield1973} (80.4 barn) indicates that the relativistic effects are small compared to other effects considered in this work. As the Born approximation becomes valid at high energy, the continuum amplitudes become effectively independent of the model potential. Therefore, the final continuum states calculated using HF and HFS converge to the same result for high electron energies. The valence amplitudes also contribute to the transition matrix element, but strongly depend on the potential model. Some minor differences in our HFS rates with those in \cite{Santra2011} , mostly confined to the third digit, can be attributed to our choice of the radial meshes which differ from those in \cite{Santra2011} and, in our case, are exponential \cite{JohnsonBook} for bound states (1000-2000 points) and log-linear \cite{Amusia1997} for continuum states (3000-25000 points) spanning from $10^{-3}/Z$ to $50 - 70$ atomic units depending on the chemical element. For these mesh parameters the results converged to those obtained on the meshes with twice the number of points.

The use of different potentials for initial and final orbitals in HF approximation results in different values of transition matrix elements in length and velocity gauges. Eq. (\ref{Photo}) is written in the length gauge, which is employed throughout this paper. It can easily be re-written in velocity gauge \cite{Grant1974} by replacing the radial integral
\begin{multline}\label{VelGauge}
\int_0^{\infty} drP_{e}(r)rP_{c}(r) \rightarrow \\
\frac{1}{\omega}\int_0^{\infty} drP_{c}(r)\left[\frac{d}{dr} + \frac{(l_e - l_c)(l_e+l_c + 1)}{2r}\right]P_{e}(r).
\end{multline}
Tab. \ref{PhotoGaugeTab} illustrates the photoionization cross-section values dependence on the gauge. The photoionization rates vary by less than 5\% and lead to an order of magnitude smaller discrepancy between length and velocity gauge results in HF potential than the discrepancy between the yields obtained with HF and HFS potentials.

\section{Fluorescence}
When the vacancy in the inner shells is created it resembles a highly excited state of the system. Such a state is unstable and quickly transitions into the lowest available energy configuration. The system can dispose of excess energy by emitting a photon (fluorescence) or an electron (Auger/Coster-Kronig processes); we first consider fluorescence.

The energy of the photon is given by $\omega_{fh} = \epsilon_f - \epsilon_h$, where $\epsilon_f$ is the energy of the initial orbital $f$ from which electron fills the hole in orbitals $h$ with energy $\epsilon_h$. This transition represents spontaneous decay and the leading order process is of the same electric dipole type as is photoionization and the same selection rules are applied. The total rate $\Gamma_F$ can be calculated using the equation
\begin{equation} \label{fluor}
\Gamma_F=\frac{4}{3}\alpha^3\omega_{fh}^3\frac{N^H_hN_f}{4l_f+2}\frac{l_>}{2l_h+1}\left|\int_0^{\infty} drP_{h}(r)rP_{f}(r)\right|^2,
\end{equation}

where $l_> = \max(l_h, l_f)$. For fluorescence calculations we used the radial orbitals obtained in a single calculation performed for the initial configuration. This approach is known to give sufficiently good accuracy for initial guess that can be further improved by means of many body perturbation theory \cite{GrantBook}, the random phase approximation \cite{Amusia1997}, or truncated configuration interaction\cite{Kozlov2013, Lepers2016}. Any of those refinements would be prohibitively expensive as we seek to evaluate all the possible electronic configurations for a given atom. The molecular effects that were ignored in this model limit the accuracy of the results obtained using atomic calculations in the molecular context. In heavy elements, fluorescence is the dominant mechanism for filling of core vacancies, while for lighter elements Auger effect defines the primary channels.

\begin{table}\center
\caption{Comparison of Auger and Coster-Kronig rates $\Gamma_A$  for atomic carbon calculated using HF and HFS potentials. Units are $10^{-3}$ a.u. A, XATOM code \cite{Santra2011}, B. Cowan code \cite{Moribayashi2008}.} 
{\renewcommand{\arraystretch}{0}%
\begin{tabular}{c c c c c c c c c}
\hline\hline
\rule{0pt}{4pt}\\
charge & configuration &  & \multicolumn{2}{c}{$KL_1L_1$} & \multicolumn{2}{c}{$KL_1L_{23}$} & \multicolumn{2}{c}{$KL_{23}L_{23}$} \\ 
\rule{0pt}{2pt}\\
& & & HF & HFS & HF & HFS & HF & HFS \\ 
\rule{0pt}{4pt}\\
\hline
\rule{0pt}{4pt}\\
+1 & $1s^12s^22p^2$ & & & & & & & \\
\rule{0pt}{2pt}\\
 & This work &  & 0.881 & 0.966 & 0.825 & 0.975 & 0.306 & 0.441 \\
 \rule{0pt}{4pt}\\
 & A & & & 0.961 & & 0.970 & & 0.439 \\
  \rule{0pt}{4pt}\\
 & B & & 0.680 & & 0.697 & & 0.392 & \\
\rule{0pt}{4pt}\\
\hline
\rule{0pt}{2pt}\\
+2 & $1s^02s^22p^2$ &  & 3.14 & 2.92 & 3.74 & 3.36 & 2.09 & 1.77 \\
\rule{0pt}{2pt}\\
 & $1s^12s^12p^2$ &  & - & - & 0.554 & 0.607 & 0.424 & 0.578 \\
\rule{0pt}{2pt}\\
 & $1s^12s^22p^1$ &  & 1.11 & 1.19 & 0.562 & 0.625 & - & - \\
\rule{0pt}{4pt}\\
\hline
\rule{0pt}{2pt}\\
+3 & $1s^02s^12p^2$ &  & - & - & 2.30 & 2.01 & 2.58 & 2.16 \\
\rule{0pt}{2pt}\\
& $1s^02s^22p^1$ &  & 3.77 & 3.51 & 2.34 & 2.02 & - & - \\
\rule{0pt}{2pt}\\
 & $1s^12s^02p^2$ &  & - & - & - & - & 0.615 & 0.626 \\
\rule{0pt}{2pt}\\
 & $1s^12s^22p^0$ &  & 1.37 & 1.41 & - & - & - & - \\
\rule{0pt}{2pt}\\
 & $1s^12s^12p^1$ &  & - & - & 0.366 & 0.376 & - & - \\
 \hline
\rule{0pt}{2pt}\\
+4 & $1s^02s^02p^2$ &  & - & - & - & - & 3.01 & 2.64\\
\rule{0pt}{2pt}\\
 & $1s^02s^22p^0$ &  & 4.46 & 3.99 & - & - & - & - \\
 \rule{0pt}{2pt}\\
 & $1s^02s^12p^1$ &  & - & - & 1.41 & 1.17 & - & - \\
\rule{0pt}{4pt}\\
\hline\hline
\end{tabular}}\label{AugerTab}
\end{table}

\section{Auger and Coster-Kronig processes}

For light atoms the dominant relaxation mechanism is the Auger effect, when the electron from a higher energy orbital $f$ fills the low energy vacancy $h$ and transfers all the excess energy to another electron $e$ from outer shell by means of Coulomb interaction. The latter electron escapes the atom if it acquires energy $E_c > 0$. By means of perturbation theory the transition rate for Auger and Coster-Kronig process is given by the expression \cite{Bhalla1971, Bhalla1973}
\begin{equation}\label{defAuger}
\Gamma_A = 2\pi\overline\sum\left|\langle h, c |\frac{1}{r_{12}}|f, i \rangle - \langle h, c |\frac{1}{r_{12}}|i, f \rangle\right|^2.
\end{equation}
Following \cite{Bhalla1973}, $\overline\sum$ stands for summation over final states and averaging over initial states. Adopting the $LS$ coupling scheme \cite{Santra2011} and summing over spin projections and magnetic quantum numbers one can write (\ref{defAuger}) as
\begin{multline}\label{Auger}
\Gamma_A = \pi\frac{N^H_hN_{if}}{2l_h+1}\sum_{l_c}\sum_{S=0}^1\sum_{L=|l_i-l_f|}^{l_i+l_f}\\
(2L+1)(2S+1)\left|M_{LS}(h, c | f, i)\right|^2,
\end{multline}
\noindent where $c$ stands for continuum orbital to which electron $i$ is promoted, $l_c$ is it's angular momentum. The weighting factor $N^H_h$ stands for the number of holes in orbital $h$, and 
\begin{equation}\nonumber
N_{if} = \left[\begin{matrix}
\frac{N_iN_f}{(4l_i+2)(4l_f+2)} & i\ne f, \\
\frac{N_i(N_i-1)}{(4l_i+1)(4l_i+2)} & i = f.
\end{matrix}\right.
\end{equation}
\noindent Applying the Wigner-Eckart theorem to matrix element in (\ref{Auger}) one obtains
\begin{widetext}
\begin{multline}
M_{LS}(h, c | f, i) = \tau(-1)^{L + l_c + l_f}\sum_K \left[ R_K(h,c,f,i) \langle l_h\parallel C^K \parallel l_f\rangle\langle l_c\parallel C^K \parallel l_i\rangle \left\{\begin{matrix}
l_h & l_c & L \\
l_i & l_f & K
\end{matrix}\right\} + \right.\\
\left.(-1)^{L+S} R_K(h,c,i,f)\langle l_h\parallel C^K \parallel l_i\rangle\langle l_c\parallel C^K \parallel l_f\rangle \left\{\begin{matrix}
l_h & l_c & L \\
l_f & l_i & K
\end{matrix}\right\}\right]
\end{multline}
\end{widetext}
\noindent where the reduced matrix element \cite{EdmondsBook}
\begin{equation}\label{Angular}
\langle l_a\parallel C^K \parallel l_b\rangle = (-1)^{l_a}\sqrt{(2l_a+1)(2l_b+1)}\left(\begin{matrix}
l_a & K & l_b \\
0 & 0 & 0
\end{matrix}\right),
\end{equation}
and the radial integral
\begin{multline}\label{Radial}
R_K(a,b,c,d) = \\
\iint_0^{\infty}d^3r_1d^3r_2 P_a(r_1) P_b(r_2) \frac{r_<^K}{r_>^{K+1}} P_c(r_1) P_d(r_2)
\end{multline}
are defined by the corresponding expressions in eq. (\ref{Angular}) and (\ref{Radial}). In the radial integral in eq. (\ref{Radial}) $r_< = \min(r_1, r_2)$ and $r_> = \max(r_1, r_2)$.

The energy of the Auger electron can become very small for certain configurations. To ensure the convergence of continuum wavefunctions in a Hartree-Fock potential one needs to integrate to some distant radial point where the asymptotic behavior is reached. Such a procedure becomes computationally expensive for low energy of Auger electrons so the cutoff energy is introduced. We used $\min(\epsilon_c) = 0.5$ a.u. as the cutoff, which we found to give sufficiently accurate results. Beyond this threshold higher order correlation effects may render the energy negative, making an excited state bound rather than a continuum function. Since we do not consider such processes in the first place it makes the cutoff procedure necessary to validate the approximations that have been adopted.Table \ref{AugerTab} lists Auger rates for all ionic states used in our simulations for carbon atom. The values for Hartree-Fock-Slater of the rates agree with those XATOM code \cite{Santra2011} to better than 1\% for a single K-shell hole, which is reduce to a maximum of 6\% for highly charged states. The difference origins from the different choice of a numerical mesh in \cite{Santra2011} and this work for both bound and continuum states. We used an increased density of points near the nucleus and the same density as \cite{Santra2011} far from the nucleus for continuum states while our bound state mesh had ten times more points than the mesh used in \cite{Santra2011}. The agreement of the rates obtained in a Hartree-Fock potential with those obtained using Cowan code \cite{Moribayashi2008}, which uses Hartree-Fock and configuration interaction (CI), is better than 25\% which is reasonable considering that we do not use a post-Hartree-Fock methods.

\section{Rate equation model for time evolution}
The accurate solution of the time-dependent Schr\"odinger equation is a challenging problem. Within perturbation theory it is convenient to use an approximation where the time-dependent wavefunctions are replaced by the time-dependent probabilities of finding the system in certain electronic configuration at a given time. This probabilities are interconnected by the processes described above and together form a system of the coupled rate equations \cite{Moribayashi2008, Santra2011}. This system can be written as
\begin{equation}\label{RateEq}
\frac{dP_i(t)}{dt} = \sum_{j \ne i} \left[\Gamma_{j\rightarrow i}(t)P_j(t) - \Gamma_{i\rightarrow j}(t)P_i(t)\right],
\end{equation} 
\noindent where $i, j$ indicate electron configurations of bound state electrons, rates $\Gamma_i$ can be both time independent, as the fluorescence and Auger processes, or time-dependent as is the case of photoionization $\Gamma_i(t) = \sigma_i I(t)$. The pulse intensity, $I(t)$, is assumed to have a Gaussian profile characterized by peak intensity and width. Although a real XFEL pulse cannot be accurately described by a single Gaussian type function, it has been shown that such an approximate description provides acceptable quantitative match to experimental dynamics \cite{Santra2007}. We ran the simulation for time intervals equal to 10 pulse widths, with the pulse centered in the middle of the time interval. A smoothing window function was also introduced at the beginning of the pulse to suppress numerical instabilities. Time intervals between each iteration of the solution were linearly decreased towards the peak of the pulse and increased afterwards. We used the implicit Newton method with precision of $10^{-6}$ for convergence of the first five starting points and a five step PECE Adams-Moulton method outwards \cite{MoultonBook}. The size of the numerical grid depends on the parameters of the pulse and indirectly on the element under consideration through the total number and magnitude of transition rates. For a model 8 keV XFEL pulse with $10$ fs width and peak intensity of $5.4 \times 10^{20}$ W/cm$^2$, 1000 points were sufficient for stable integration for light atoms such as C, N, and O, but the required density of the numerical grid grows rapidly for heavier elements. For sulfur we used  20000 points, and for iron $10^6$ points were required. This is to be expected for heavier atoms for which the photoionization cross-section for K-shell grows as $Z^6$ \cite{BetheSalpeterBook} which requires a much smaller time step for any finite difference scheme. The calculation time is further increased by the rapid growth of the number of available decay channels for each of $N$ possible configurations which can be evaluated as  
\begin{equation}
N = \prod_{a = 1}^{n_o} (q_a + 1),
\end{equation}
where $n_o$ is the number of orbitals, $q_a$ is the number of electrons in the initial orbital $a$ that can be ionized by photons of a given frequency. It may be noticed that for some pulses and moderately heavy atoms the K shell ionization potential may drop below the photon energy due to the reduction in nuclear charge screening as the atom looses valence electrons during the pulse. This effect was discussed for the neon atom \cite{Young2010} and is demonstrated below for an iron atom.

\section{Charge state dynamics}
When a single molecule or a cluster is exposed to a short bright pulse of an XFEL it builds up substantial net charge during the pulse. Therefore the electronic structure of a target is quickly distorted from its original configuration. The standard approach to such complex dynamics is to treat atoms as independent subject to external perturbations. We solved the rate equations using the rates calculated with HF and HFS potentials for carbon, nitrogen, oxygen, sulfur, and iron atoms. The first three atoms are relatively light. Together they represent the majority of the atom types encountered in biological molecules. Sulfur and iron, on the other hand, are relatively heavy. Although usually present in small amounts in proteins, their physical properties with regards to the interaction with X-rays are starkly different to those of the light elements. 

\begin{table}\center
\caption{Probabilities (in $\%$) of various ionization stages for C, N, O, S, and Fe atoms at the end of the 8 keV $10$fs XFEL pulse with peak intensity $5.4\times 10^{20}$ W/cm$^2$. Only the ions with yield greater than $0.5\%$ are listed. Initial states are taken to be $100\%$ of a corresponding neutral atom.} 
{\renewcommand{\arraystretch}{0}%
\begin{tabular}{c c c c c c c c c}
\hline\hline
\rule{0pt}{4pt}\\
element & \multicolumn{8}{c}{total ionic charge} \\ 
\rule{0pt}{2pt}\\
& \multicolumn{2}{c}{$Z-3$} & \multicolumn{2}{c}{$Z-2$} & \multicolumn{2}{c}{$Z-1$} & \multicolumn{2}{c}{$Z$} \\ 
\rule{0pt}{2pt}\\
& HF & HFS & HF & HFS & HF & HFS & HF & HFS \\ 
\rule{0pt}{4pt}\\
\hline
\rule{0pt}{4pt}\\
C & 1.1 & 1.2 & 8.6 & 10.8 & 53.7 & 47.7 & 36.5 & 40.3 \\
\rule{0pt}{4pt}\\
\hline
\rule{0pt}{4pt}\\
N & - & - & 5.8 & 6.0 & 57.9 & 62.9 & 27.5 & 30.7 \\
\rule{0pt}{4pt}\\
\hline
\rule{0pt}{4pt}\\
O & - & - & 1.4 & 1.7 & 42.4 & 43.0 & 56.1 & 55.1 \\
\rule{0pt}{4pt}\\
\hline
\rule{0pt}{4pt}\\
S & - & - & - & - & 9.8 & 8.8 & 90.9 & 91.2 \\
\rule{0pt}{4pt}\\
\hline
\rule{0pt}{4pt}\\
Fe & 8.0 & 6.3 & 66.0 & 59.9 & 25.6 & 32.6 & - & 1.2 \\
\rule{0pt}{4pt}\\
\hline\hline
\end{tabular}}\label{IonYield}
\end{table}

Tab. \ref{IonYield} represents the results for ion yields obtained by solving eq. (\ref{RateEq}) for charge state dynamics for carbon, nitrogen, and oxygen atoms in the field of a $10$ fs XFEL pulse with peak intensity $5.4\times 10^{20}$ W/cm$^2$ focused in a $100\times100$ nm$^2$ area. For light elements, only three charge states have an appreciable population by the end of the pulse: complete ionization ($C^{6+}$,$N^{7+}$,$O^{8+}$), one electron retained ($C^{5+}$,$N^{6+}$,$O^{7+}$), and two electrons retained ($C^{4+}$,$N^{5+}$,$O^{6+}$). For all of these elements the most likely state is that with one remaining electron. The HF and HFS approximations produce similar results that differ by no more than $6\%$. This result changes by less than 1\% when a velocity gauge for photoionization matrix elements generated using HF potential was used. This leads to the conclusion that the difference in Auger rates that are unaffected by the choice of the gauge has much more influence on the ion yield than the photoionization and fluorescence rates for light elements. 

\begin{figure}[t]
\center{\includegraphics[width=1\linewidth]{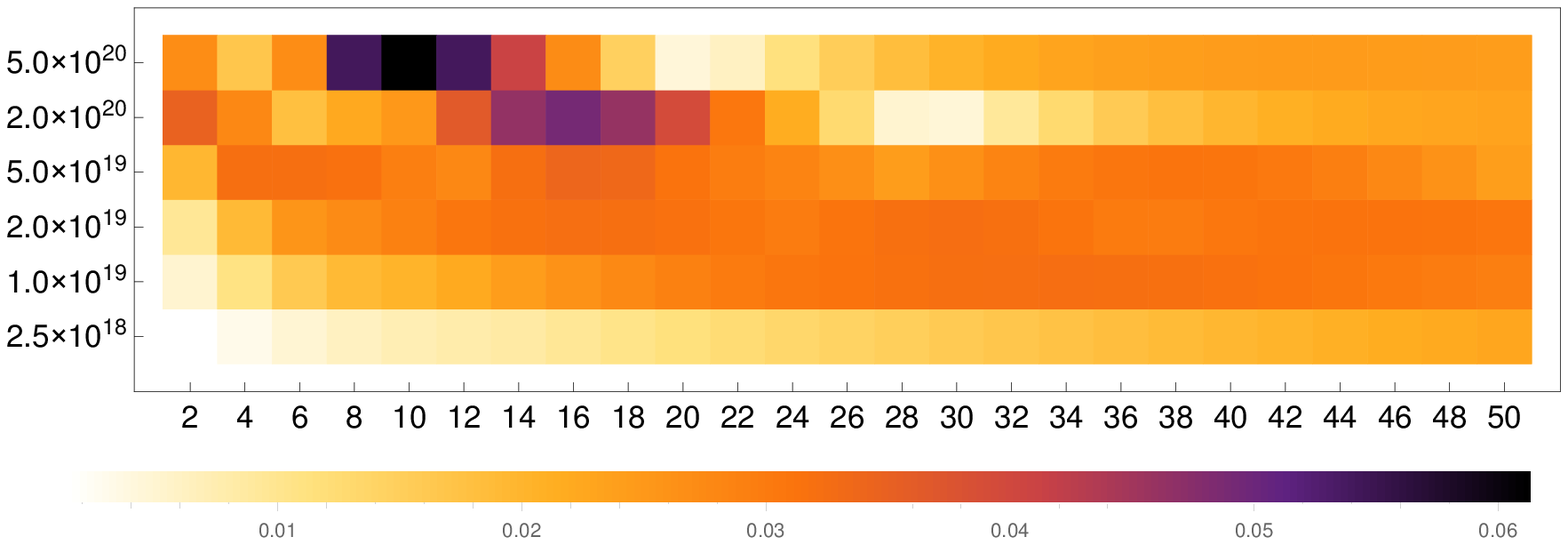}}
\center{\includegraphics[width=1\linewidth]{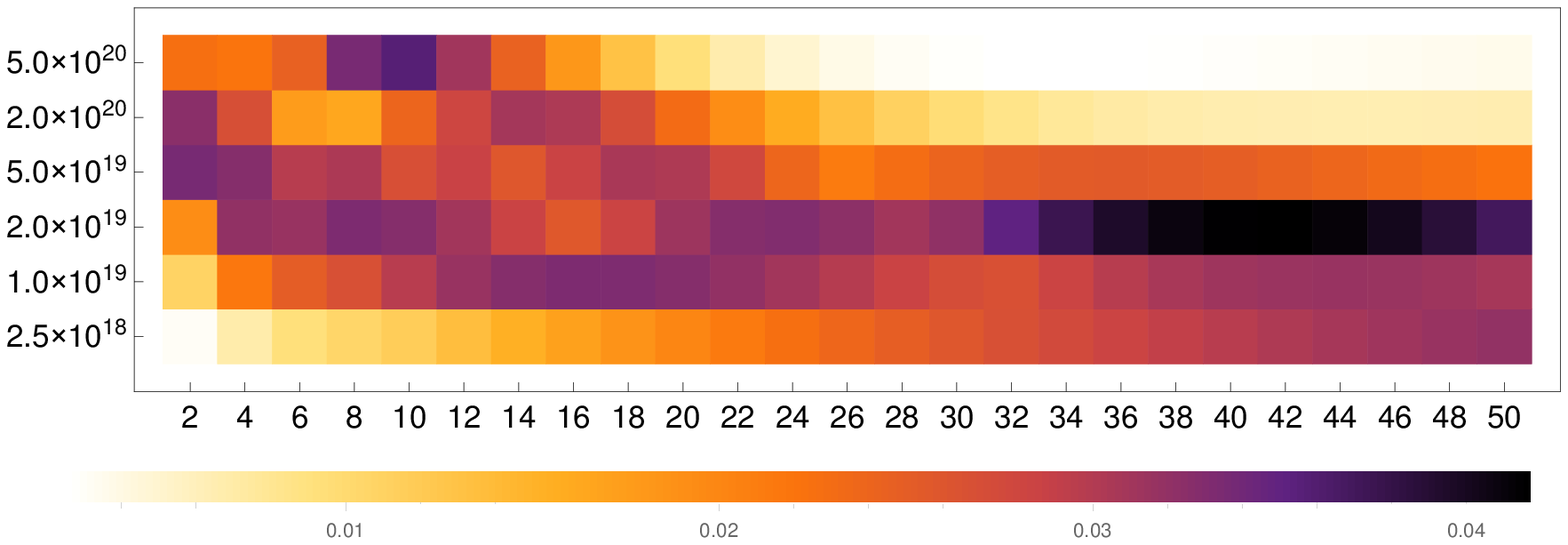}}
\center{\includegraphics[width=1\linewidth]{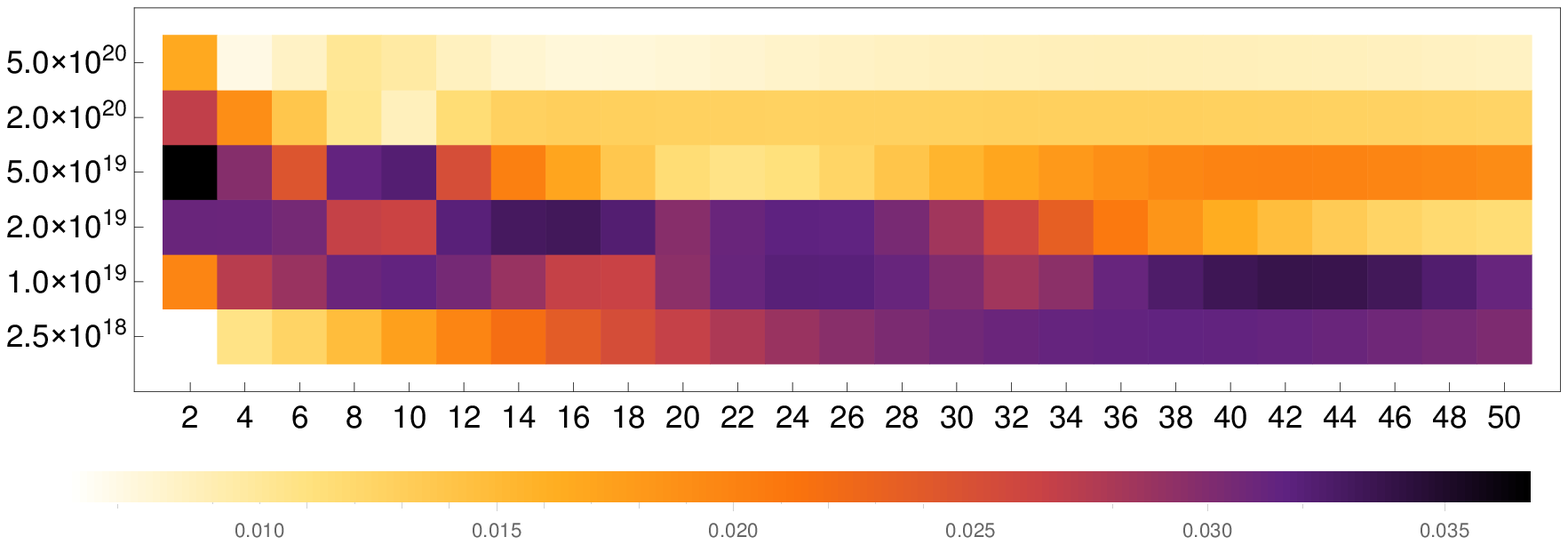}}
\center{\includegraphics[width=1\linewidth]{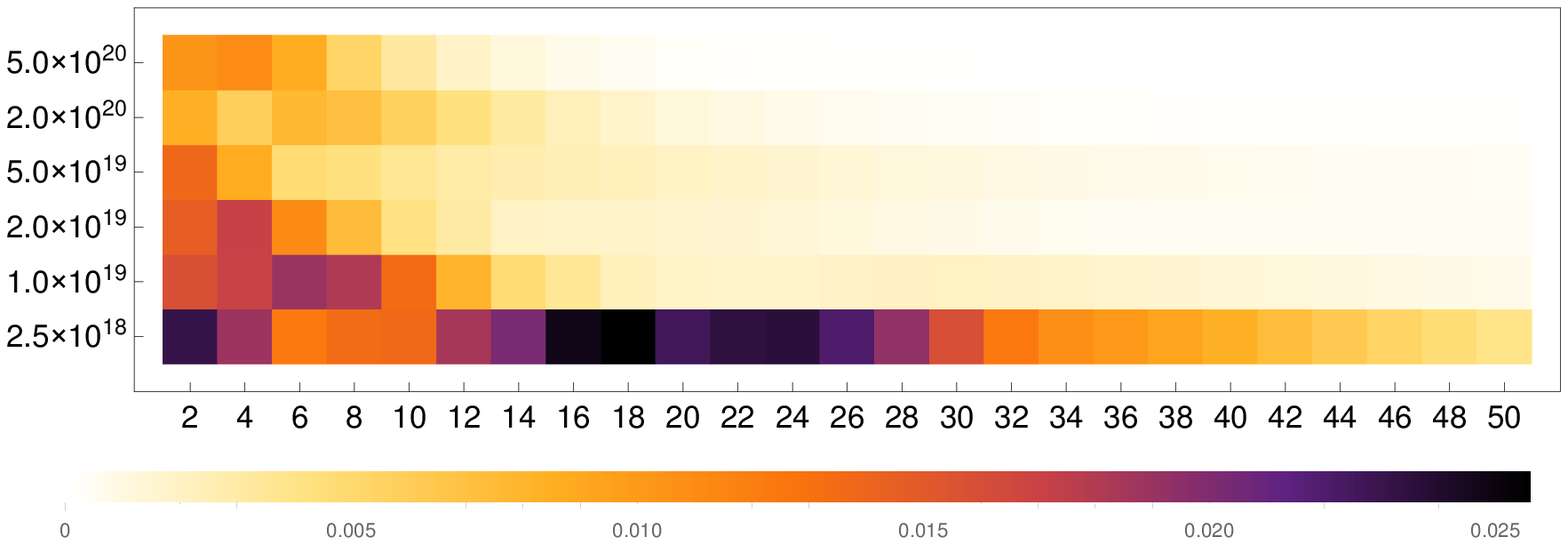}}
\center{\includegraphics[width=1\linewidth]{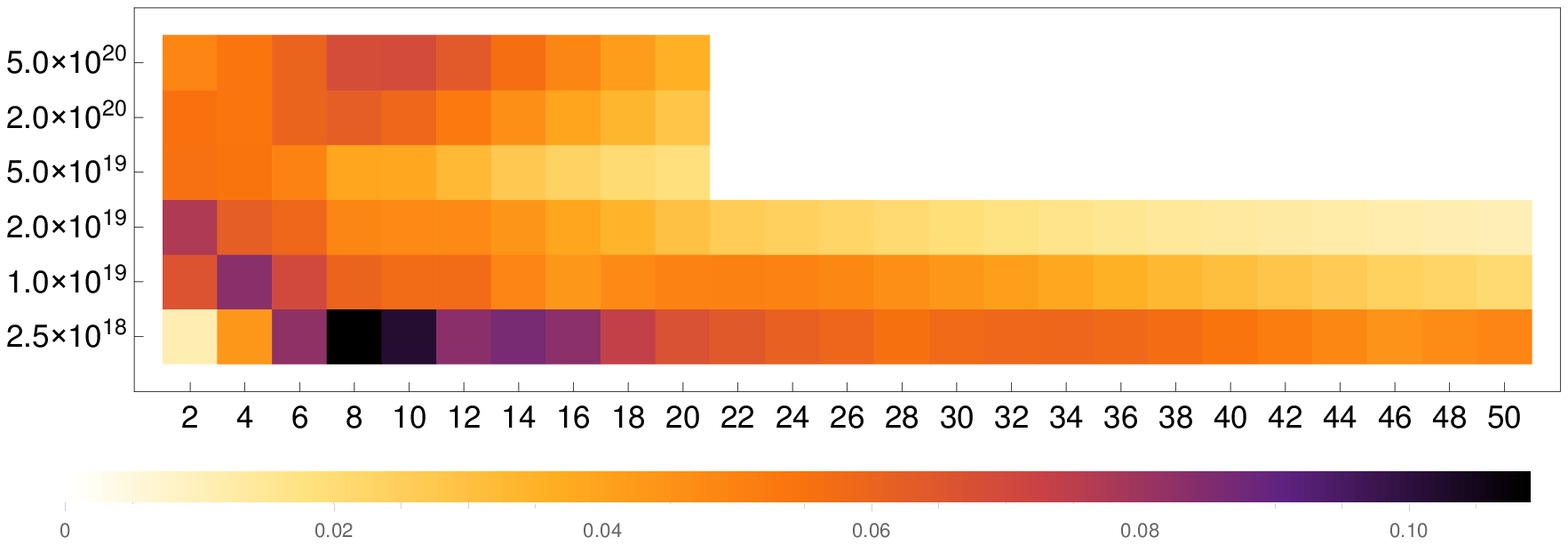}}
\caption{Maximum discrepancy in predicted ion yield by HF and HFS models as a function of the XFEL pulse width and peak intensity. From top to bottom: C, N, O, S, Fe; abscissa represents the pulse width in femtoseconds, ordinate represents the peak intensity in W/cm$^2$.}\label{TileChart}
\label{LightHF}
\end{figure}

\begin{figure*}[t]
\begin{minipage}[t]{0.49\linewidth}
\center{\includegraphics[width=1\linewidth]{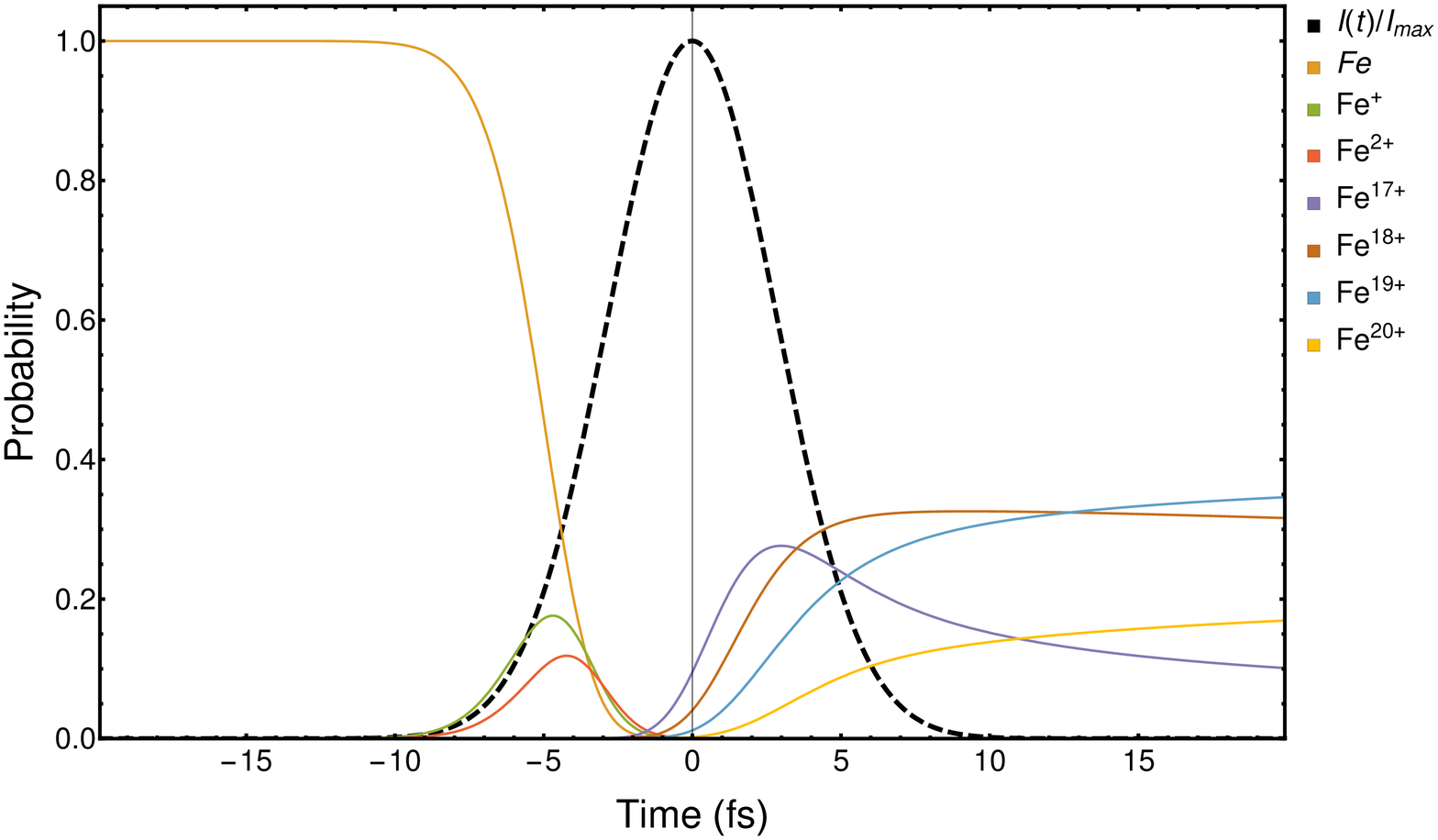}}
\end{minipage}
\hfill
\begin{minipage}[t]{0.49\linewidth}
\center{\includegraphics[width=1\linewidth]{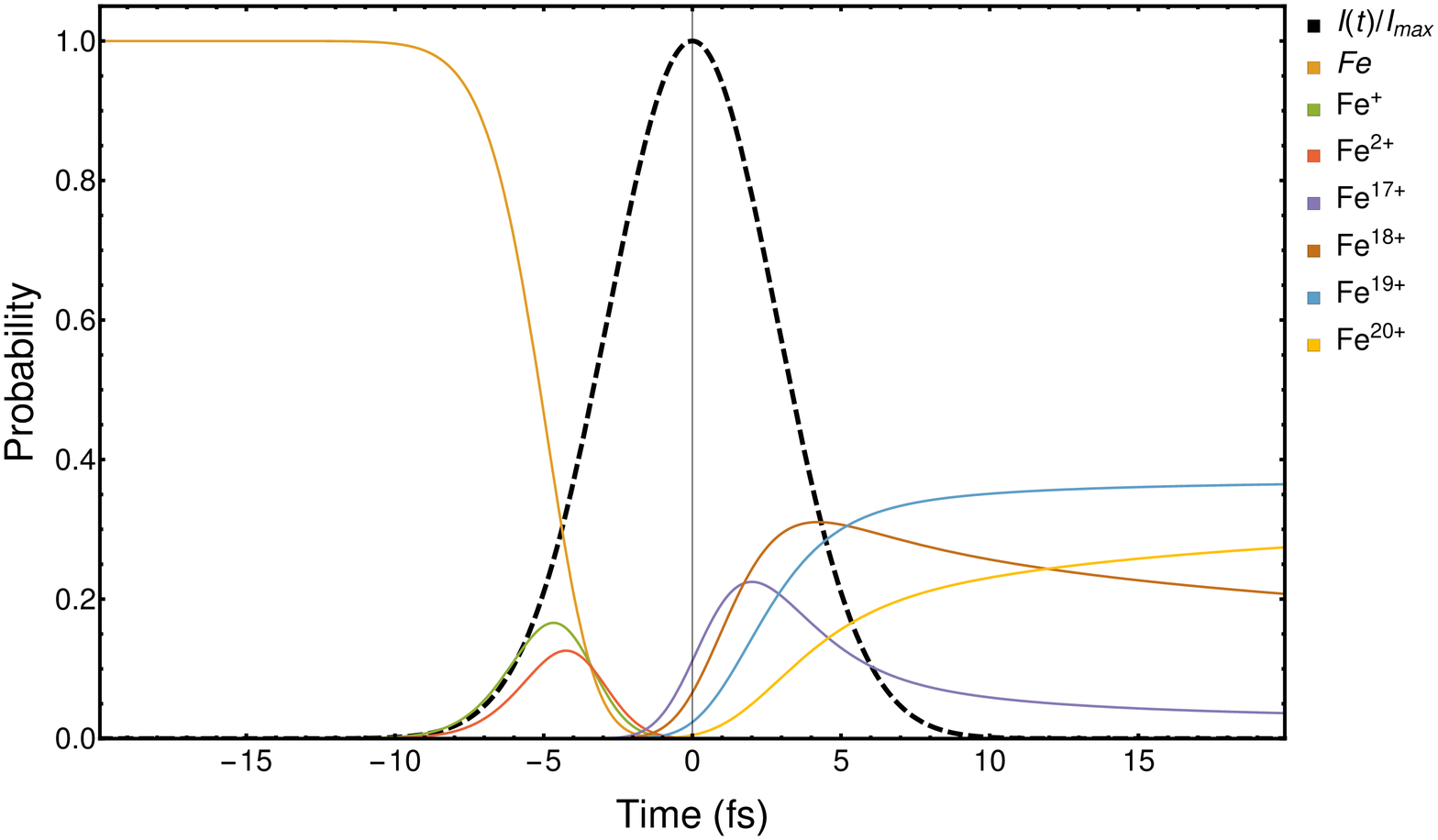}}
\end{minipage}
\caption{Total charge states of iron atom calculated in HF (left) and HFS (right) approximations. The peak intensity is $2.5\times 10^{18}$ W/cm$^2$ and the pulse width is 8 fs.}\label{Heavy}
\end{figure*}

To compare the predictions of HFS and HF approximations one can use the maximum difference in predicted ion yields. For example from Tab. \ref{IonYield} one can conclude that for the given pulse parameters two methods predict a maximum of 6\% difference for carbon (C$^{5+}$) and 7\% for iron (Fe$^{25+}$). Fig. \ref{TileChart} depicts this discrepancy measure as a function of the peak intensity and the width of the XFEL pulse. The shape of the pulse is assumed to be Gaussian in all cases. One can see that the maximum discrepancy for light elements reaches about 6\% for carbon atoms in the given range of pulse parameters. The maximum discrepancy shows a downward trend with increase of atomic number for the light elements. For sulfur it reaches the maximum of $2.5$\% for the peak intensity of $2.5\times 10^{18}$ W/cm$^2$ and the pulse width of 16 fs (see Fig. \ref{Heavy}). For the iron atom we observed the largest discrepancy of all the elements. It reaches $10.9$\% for the pulse with the peak intensity of $2.5\times 10^{18}$ W/cm$^2$ and the width of 8 fs. The reduction in maximum discrepancy between the two approaches was observed for longer pulses with higher intensities. For iron we haven't carried out the simulations for pulses with the peak intensities beyond $5\times 10^{19}$ W/cm$^2$ and for widths greater than 22 fs. For the pulse parameters beyond these values the time required for each simulation exceeded four hours. A further convergence of both approaches is expected in this region due to the large radiation dose delivered to the sample.

In our HF calculation the K-shell ionization potential for neutral iron atom is $261.4$ a.u. which is comparable to the $294.1$ a.u. photon energy in the model XFEL pulse. For nitrogen-like iron the K shell ionization potential rises to $296.1$ a.u., therefore blocking photoionization in that shell. A similar effect can take place in the system described in \cite{Ilme2014} where the simulations \cite{Caleman2010} showed a small difference in the ion yields of iron atoms irradiated with XFEL pulses with photon energy slightly above (7.36 keV) and below (6.86 keV) the neutral K-shell ionization potentials. For the pulse parameters considered in this work our simulations yielded almost complete liberation of all electrons except those occupying K-shell for both photon energies. 

To illustrate the effect of dynamics on the ion yield we considered two pairs of pulses, the first pair with photon energies of 6.86 keV and 7.36 keV, and the second pair with photon energies of 7.89 keV and 8.43 keV respectively. The first pair represents the same photon energies as those used in \cite{Ilme2014} which are slightly below and above the K-shell ionization threshold of neutral iron. The second pair is chosen such that the photon energies are below and above the K-shell ionization threshold of neon-like iron. In both cases the pulses had Gaussian profile with the width of $20$ fs and  $1.0\times 10^{19}$ W/cm$^2$ peak intensity. The results of the simulations are summarized in tab. \ref{IronYield}. One can see that there is very little difference between the ion yields for the first pair of pulses. This occurs because after a few ionization events the K-shell threshold drops below 7.36 keV, blocking the K-shell ionization. This effect doesn't happen in the second pair of pulses until the ion reaches a highly charged state, such as that of the neon-like iron. When that happens, the first beam (7.89 keV) stops ionizing the K-shell, while the second beam remains able to do so for about four more ionization events. Considering the relatively small number of remaining electrons at that stage, the effect of the ionization dynamics becomes more pronounced for the second pair of pulses than for the first one.

\begin{table}\center
\caption{Ion yields (in $\%$) of iron atom in XFEL pulses with photon energy 6.86 keV and 7.36 keV used in \cite{Ilme2014}. Pulse parameters: $16$fs width, $1.0\times 10^{19}$ W/cm$^2$ peak intensity. Only the ions with yield greater than $3\%$ are listed.} 
{\renewcommand{\arraystretch}{0}%
\begin{tabular}{c c c c c}
\hline\hline
\rule{0pt}{4pt}\\
charge & 6.86 keV & 7.36 keV & 7.89 keV & 8.43 keV\\ 
\rule{0pt}{4pt}\\
\hline
\rule{0pt}{4pt}\\
$23^+$ & - & - & - & 14.1\\
\rule{0pt}{2pt}\\
$22^+$ & 23.3 & 23.1 & 29.8 & 45.9\\
\rule{0pt}{2pt}\\
$21^+$ & 40.7 & 40.3 & 43.6 & 34.2\\
\rule{0pt}{2pt}\\
$20^+$ & 24.1 & 24.8 & 20.6 & 5.5\\
\rule{0pt}{2pt}\\
$19^+$ & 8.9 & 9.1 & 5.0 & -\\
\rule{0pt}{4pt}\\
\hline\hline
\end{tabular}}\label{IronYield}
\end{table}

\section{Conclusions}

We carried out simulations of radiation damage dynamics for light and moderately heavy atoms of biological significance using Hartree-Fock and Hartree-Fock-Slater potentials. Comparing the yields of different ionization states we observed that for high peak intensities and long pulses both approximations are in good agreement.  The convergence of the two approaches is expected on the basis on the perturbation theory. Indeed when the radiation dose absorbed by an atom exceeds the total ionization potential, the details of the atomic structure become increasingly less relevant for the ionization dynamics. By comparing the predictions of photoionization cross-sections with an existing relativistic Hartree-Fock-Slater calculations confirm that the relativistic effects make a small contribution to the ion yields compared to the exchange potential. A larger, but still small effect was observed when comparing the length and velocity gauges of the electric dipole matrix element in photoionization. The variation of ion yields obtained by comparing the results of HF calculations with length and velocity gauges was about an order of magnitude smaller than the variation obtained when comparing HF and HFS calculations in length gauge.

Among the light elements the largest discrepancy in ion yields was 6\% for carbon atom in the 10 femtosecond pulse with the peak intensity of $5.0\times 10^{20}$ W/cm$^2$. The best agreement was observed for sulfur atom, where the discrepancy in ion yields at no point exceeded 3\%. For the iron atom we observed the largest discrepancy among the elements considered which is about 11\% for HF and HFS results obtained for 8 femtosecond pulse with peak intensity of $2.5\times 10^{18}$ W/cm$^2$. The reason for the discrepancy reaching the maximum at the low end of the XFEL pulse parameters may be found in the relatively strong susceptibility of iron to the incoming radiation compared to the lighter elements. Indeed K and L shells of iron are very potent absorbers of 8 keV X-ray photons, while light elements like carbon, oxygen, and nitrogen experience rather small ionization in such ``weak" pulse.  

To illustrate the sensitivity of the ionization dynamics to the photon energy we carried out an additional set of simulations for an iron atom subject to the two sets of pulses. The first set had a photon energies above and below of a neutral K shell threshold (6.86 keV and 7.36 keV) while in the second set included the photon energies above and below a neon-like iron K shell ionization potential (7.89 keV and 8.43 keV). While the first set of the pulses produced a very similar ion yields, the second set predicted a considerable variation in the yields. In both sets there is an ionization stage when only the higher energy pulse can ionize K-shell before its ionization potential becomes larger than the photon energy. But for the second set this ionization stage is reached at a much later time, where it affects the resultant ion yield in a much bigger way, as is seen from the simulation results.

To summarize, both Hartree-Fock and Hartree-Fock-Slater potentials are equally viable for simulations of interactions of biomolecules in the field of an XFEL pulse. The differences between both approaches emerges for a shorter and less intense XFEL pulses in heavier elements and short and more intense pulses for light elements. The ion yields, cross-sections, and decay rates produced by two approximations can be used as an input for molecular dynamics simulations to estimate errors when comparing theoretical results with experiment \cite{Ostlin2017, Ho2017}. The gauge choice has a small effect on the ion yield for the elements considered in this work, although it may be much larger if a heavier elements are present in the sample. While negligible for light elements, relativistic effects substantially affect cross-sections and decay rates in heavy elements, where j-j coupling scheme might be more preferable to LS scheme used in this work. The impact of relativistic effects on the ion yields for moderately heavy elements such as iron are significant and will be a subject of future work.

It should be noted that a similar approach can be used to investigate the problem of highly charged ion production during laser ablation \cite{Hassanein2015}. This can be achieved by adding multiphoton absorption channels and taking into account resonance effects in photon absorption as it was done for XUV XFEL modeling \cite{Makris2009}. Another prospective avenue to investigate is the energy deposition into a surface from collisions with highly charged ion beams \cite{Sosolik2015}. Although the surface exposure time is usually well beyond the time scale of our model, the charge transfer occurs on the femtosecond time scale and therefore can be modeled using a rate equation approach.

\section*{Acknowledgements}
The authors acknowledge the support of the Australian Research Council through the Centre of Excellence for Advanced Molecular Imaging.

\end{document}